\documentclass{llncs}

\usepackage{amsmath}
\usepackage{amssymb}
\usepackage{latexsym}
\usepackage{graphicx}

\usepackage{mdwmath}
\usepackage{mdwtab}
\usepackage[tight,footnotesize]{subfigure}

\newcommand\bcdot{\ensuremath{
  \mathchoice
   {\mskip\thinmuskip\lower0.2ex\hbox{\scalebox{1.5}{$\cdot$}}\mskip\thinmuskip}}
   {\mskip\thinmuskip\lower0.2ex\hbox{\scalebox{1.5}{$\cdot$}}\mskip\thinmuskip}
   {\lower0.3ex\hbox{\scalebox{1.2}{$\cdot$}}}
   {\lower0.3ex\hbox{\scalebox{1.2}{$\cdot$}}}
}

\begin{document}

\title{Dependability-Explicit Engineering with Event-B:  Overview of Recent Achievements}

\titlerunning{Dependability-Explicit Engineering with Event-B}

\author{Elena Troubitsyna}

\authorrunning{Elena Troubitsyna}

\institute{\AA{}bo Akademi University, Turku, Finland \\
\email{\textbraceleft  Elena.Troubitsyna\textbraceright@abo.fi}}

\maketitle

\begin{abstract}
Event-B has been actively used within the EU Deploy project  to model dependable systems from  various application domains.  As a result, we have created a number of formal approaches to explicitly reason about dependability in the refinement process. In this paper we overview the work on formal engineering of dependable systems carried out in the Deploy project. We outline our approaches to integrating safety analysis into the development process, modelling fault tolerant systems and probabilistic dependability evaluation. We discuss achievements and challenges in development of dependable systems within the  Event-B framework.

\end{abstract}

\section{Introduction}
Nowadays we tend to place increasing reliance on computer-based systems and software which they are running. The degree of reliance that we can justifiably place on a system is expressed by the notion of \emph{dependability}~\cite{ALR00}. However, the analysis of recent software-caused accidents has shown that the current development process is inadequate for achieving high degree of dependability. While a number of existing methods and tools address certain aspects of dependable systems development, there is still a lack of a general viable \emph{dependability-explicit} techniques for developing software for complex systems. 

To address this issue, in the FP7 EU Deploy project~\cite{DEPLOY} we have proposed a number of approaches that allow the designers to explicitly address dependability throughout the entire system development by refinement in Event-B~\cite{EventB}. In this paper we briefly overview the approaches that have been mainly proposed by the researchers from \AA{}bo Akademi University. The goal of this paper is to present some evidences that Event-B consitutes a suitable framework for formal dependability-explicit development.

\section{Engineering Dependable Systems with Event-B}

\subsection{Event-B} 
Currently, complexity is perceived as the main threat to dependability. To cope with the system complexity, we need scalable formal techniques to explicitly address various dependability
aspects throughout the entire development cycle. It is widely recognised that system complexity can be managed via abstract modelling, decomposition and iterative development. Event-B~\cite{EventB}  is a formal top-down development approach to correct-by-construction system development. Development in Event-B starts from defining a high-level specification that represents the system behavior and properties in a highly abstract way. The main development technique -- refinement -- allows us to ensure that a concrete specification preserves the globally observable behaviour and properties of the abstract specification, i.e., verify  correctness with respect to the abstract specification. Verification of each refinement step is done by proofs. The Rodin platform~\cite{RODINPLAT} automates modelling and verification in Event-B.

\subsection{Dependability in System Development}
The notion of dependability encompasses a wide range of system properties. Traditionally,  dependability can by
characterised by such attributes~\cite{ALR00} as reliability, safety, availability, maintainability, confidentiality and integrity. In the Deploy project, the main focus has been on developing techniques addressing safety, reliability and availability.

The system dependability is impaired by failures, errors and faults~\cite{ALR00}. To break the chain of propagation of a fault --  a physical defect or malfunction of a system component -- towards the system boundary, the system designers employ  a variety of techniques  to avoid and remove faults, as well as tolerate and forecast them. Let us now discuss the ways in which Event-B facilitates development of dependable systems. 

The main purpose of \emph{fault prevention }(or fault avoidance) techniques is
to avoid occurrence or introduction of faults during the development process. Development in Event-B allows the designers to better understand the system requirements and properties and express them in precise mathematical way. The verification that proceeds hand-in-hand with the modelling enables early identification of design errors and avoid dependability-impairing failures.

\emph{Fault tolerance} methods are used to design a system in such a way that
it is capable of functioning despite the presence of faults. While specifying fault tolerant systems in Event-B, we model not only nominal system behaviour but also failure occurrence and fault tolerance as an intrinsic part of the system specification. It allows us to formally underpin fault assumptions and rigorously define fault tolerance mechanisms. 

\emph{Fault removal }is a set of techniques for identifying and removing the
causes of errors. The fault
removal process at the development stage starts with system verification,
which is followed by diagnosis and correction steps. While modelling systems in Event-B,  we rely on proofs, probabilistic extension of Event-B and associated probabilistic model checking to verify correctness of functional behaviour and satisfaction of the desired dependability attributes.

\emph{Fault forecasting} aims at evaluation of the impact of fault occurrence and
activation on the system behaviour. Such an evaluation has qualitative and
quantitative aspects. The qualitative analysis helps to designate and classify
failure modes as well as identify combinations of faults of components that
may potentially lead to a system failure. We have demonstrated that how a seamless integration between Event-B and  various techniques for safety analysis facilitate qualitative assessment  of the impact of faults on the system dependability. The probabilistic extension of Event-B allows for the quantitative 
  assessment of to what extent certain attributes of dependability are satisfied.

Therefore, we believe that  Event-B constitutes a suitable and versatile framework for creating a rigorous dependability-explicit development process.  Next we overview in a more details our contributions to attaining establishing dependability-explicit development process with Event-B.

\section{Formal development of fault tolerant mode-rich systems}
A widely used mechanism for achieving fault tolerance is based on the notion of modes. In our work~\cite{FMICS10,Safecomp10,SCP12},  we have proposed an approach to formal development of fault tolerant mode-rich systems.  Such systems achieve fault tolerance by rollbacking to specific degraded modes. The proposed formal development process allows the designers to develop a system in a layered fashion. Essentially, it consists of a number of steps gradually unfolding system architectural layers by refinement. Moreover, we prove the consistency between mode transitions on adjacent architectural layers, which significantly improves scalability of verification.
It has been noted that testing and model checking of the systems with complex mode transition schemes suffers from poor scalability. We have overcame this problem by relying on  incremental verification of global mode consistency properties by proof and hence achieved a good scalability.

In our approach to modelling mode-rich systems~\cite{FMICS10,Safecomp10,SCP12}, we have focused on verification of consistency of a predefined mode logic. In~\cite{APSEC11}, we have proposed to conduct Failure Modes and Effects Analysis (FMEA) of each operational mode to identify mode transitions required to implement fault tolerance. Fault tolerance is achieved by two different means -- transitions to a more degraded mode and dynamic reconfiguration using redundant components. Furthermore,  we have investigated a complex interplay between the states of components during reconfiguration and the system modes.

\section{Goal-oriented refinement of reconfigurable systems}
In~\cite{IGI11,EDCC12,SERENEIn12}, we have investigated the problem of ensuring safety and fault tolerance of mobile agent systems. The work has resulted in  defining the modelling patterns to represent agent roles in dynamic scopes and deriving the logical conditions to ensure system dependability.

In~\cite{ADA12}, we have continued our study of multi-agent systems and have proposed a goal-oriented approach to development of multi-agent systems.
It is currently recognized that the goal-oriented development facilitates design of complex dynamically adaptable systems. In goal-oriented development the system requirements are defined in terms of goals -- the functional and non-functional objectives that a system should achieve. Often changes in system operational environment, e.g., caused by failures of agents -- independent system components of various types -- might hinder achieving the desired goals.
In [ADA] we have proposed a formal development approach that ensures goal reachability "by construction".
Essentially, our approach allows the developers to define system goals at different levels of abstraction and guarantee goal reachability despite agent failures. We have derived refinement patterns modelling the mechanisms for dynamic system reconfiguration by  reallocating goals from failed agents to healthy ones and, per se, guarantee dependability.
We believe that our approach offers a scalable technique for formal development of dynamically reconfigurable dependable systems.

While refining a reconfigurable system, we had to assume that sufficient amount of agents would remain operational to achieve the desired goals. In \cite{SAFECOMP12}, we have demonstrated how to integrate probabilistic analysis to quantitatively assess the likelihood of goal reachability despite failures. The rigorous refinement process has allowed us to establish the precise relationships between component failures and goal reachability. We have assessed the derived reconfigurable  system architecture to quantitatively verify that it achieves the desired reliability  and performance objectives. This was accomplished by relying on the probabilistic extension of Event-B to verify reliability and performance properties using PRISM model checker~\cite{PRISM}.

\section{Integrating Safety Analysis into Formal Development}
In~\cite{HASE11}, we have demonstrated how to combine formal modelling and refinement with Failure Modes and Effects Analysis (FMEA). We have defined a set of patterns formalising the requirements derived from FMEA as well as automated their integration into the formal specification. The proposed approach facilitates formal development and improves traceability of safety requirements.
The approach proposed in this paper allows us to automate the formal development process via two main steps: choice of suitable patterns that generically define FMEA result, and instantiation of chosen patterns with model-specific information.  Our approach allows the developers to verify (by proofs) that safety invariants are preserved in spite of identified component failure modes. Hence we believe that it provides a useful support for formal development and improves traceability of safety requirements.

The use of an evidence generated from formal analysis is still an open issue in the system certification process. Sometimes the formal proofs
deemed to be too complex and cause doubts regarding their trustworthiness  as the evidence in safety cases of safety-critical systems. Another open issue related to the formal modelling process is whether the
obtained formal model adequately represents safety requirements described in a
system  specification. In our work\cite{SERENEYu12}  we proposed an approach to linking formal modelling in Event-B  with safety cases. We give the classification of safety requirements and
define how each class can be represented in a formal specification. The approach allows the developers to obtain a consistent system
specification that facilitate deriving a "sufficient" safety case.

The systems, whose components are susceptible to various kinds of faults, never are  "absolutely" safe, i.e.,  certain combinations of failures may lead to an occurrence of a hazard  -- a potentially dangerous situation breaching safety requirements. To demonstrate that the probability of a hazard occurrence is acceptably low, in~\cite{SERENE11} we have presented a formal approach to integrating quantitative safety analysis into formal system development by refinement in Event-B. Essentially, our approach can be seen as a process of extracting a fault tree -- a logical representation of a hazardous situation in terms of the primitives
used at different abstraction layers. Eventually, we arrive at the representation
of a hazard in terms of the failures of basic system components, which allows us to calculate probability of a hazard occurrence.
The proposed approach is based on a probabilistic extension of Event-B~\cite{IFM12}. It enables development of systems that are not only correct but also safe by construction.

\section{Quantitative Assessment of Dependability}
To facilitate dependability-explicit development in the probabilistic Event-B~\cite{IFM12}, we strengthened the notion of Event-B refinement by requiring that a refined
model, besides being a proper functional refinement of its more abstract
counterpart, also satisfies a number of quantitative constraints. These constraints
ensure that the refined model improves (or at least preserves) the current probabilistic measures of system dependability attributes. In our
work, these additional constraints are usually derived from the fundamental
properties of Markov processes.
To validate the proposed approaches, in Deploy we have conducted a number of
case studies  including  formal development and quantitative assessment
of a fault tolerant satellite system, formal modelling
integrated with safety analysis of a radio-based railway crossing controller, service-oriented system etc. This work allows the designers to to evaluate the impact of
the chosen design decisions on system dependability.

\section{Discussion}
Our work on formal engineering of dependable systems in the EU Deploy project  has resulted in two types of approaches:
\begin{itemize}
\item the approaches that focus on creating modelling patterns and guidelines for representing and verifying certain resilience-related behavior
\item	the approaches that  integrate (external) techniques for safety and reliability analysis into the formal development process of Event-B.
\end{itemize}
A tight cooperation with the Deploy industrial partners has allowed us to gain rich experience in modelling dependable systems from the transportation, aerospace and business information system domains. The development of industrial-scale systems has emphasized the need for scalability in formal modelling and automatic tool support. It has fostered the research on modularisation and decomposition techniques for Event-B as well as development of various plug-ins.  Moreover, it has led to understanding importance  of heterogenous modelling techniques to address a variety of dependability aspects.

In general, we believe that Event-B offers a powerful formal technique for engineering dependable systems.  To leverage scalability and industrial relevance of the method, we will continue to enlarge the set of modelling patterns for representing various dependability aspects, strengthening automatic tool support and enriching its capabilities via dedicated plug-ins to the Rodin platform.

\bibliographystyle{plain}

\end{document}